\documentclass[aps,twocolumn,preprintnumbers]{revtex4}

\usepackage{graphicx}  
\usepackage{subfigure}
\usepackage{multirow}

\usepackage{fancyhdr}
\usepackage{longtable}
\usepackage{parskip}
\usepackage[T1]{fontenc}
\usepackage{dcolumn}   

\usepackage{bm}        
\usepackage{amsfonts}  
\usepackage{amsmath}   
\usepackage{amssymb}   

\usepackage{mathrsfs}
\usepackage{mathtools}

\usepackage{slashed}
\usepackage{color}
\usepackage{xcolor}
\usepackage{appendix}

\usepackage[colorlinks,linkcolor=red,anchorcolor=blue,citecolor=blue]{hyperref}
\usepackage{simplewick}
\usepackage{tikz}

\begin{document}
\title{Identifying Hadronic Molecular States with a Neural Network}
\author{Chang Chen $^{1}$}
\email{1901110078@pku.edu.cn}
\author{Hao Chen $^{1}$}
\email{haochen0393@pku.edu.cn}
\author{Wen-Qi Niu$^{1}$}
\email{1701110076@pku.edu.cn}
\author{Han-Qing Zheng$^{2}$}
\email{zhenghq@scu.edu.cn}

\affiliation{$^{1}$\it Department of Physics and State Key Laboratory of Nuclear Physics and Technology,\\
Peking University, Beijing 100871, China}

\affiliation{$^{2}$\it College of Physics, Sichuan University, Chengdu, Sichuan 610065, China}

\date{\today}

\begin{abstract}
Neural networks are trained to judge whether or not an exotic state is a hadronic molecule of a given channel according its line-shapes. This method performs well in both trainings and validation tests. As applications, it is applied to study $X(3872)$, $X(4260)$ and $Z_c(3900)$. The results show that $Z_c(3900)$ should be regarded as a $\bar{D}^* D$ molecular state but $X(3872)$ not. As for $X(4260)$, it can not be a molecular state of $\chi_{c0}\omega$. Some discussions on $X_1(2900)$ are also provided.
\end{abstract}

\maketitle

\section{Introduction}
Exotic hadron states, refer to those hadron states that do not appear to fit with the expectations for an ordinary $q\bar{q}$ or $qqq$ hadron in the quark model. A situation is that an exotic state may have multiquark constituents, e.g., its valence quarks are $q\bar{q}q\bar{q}$ or $qq\bar{q}\bar{q}\bar{q}$. One explanation is that these states are compact or "elementary" states, where quarks and anti-quarks are building blocks, they form a compact core and interact with each other by exchanging gluons. However, since these exotic states usually appear near the two-hadron thresholds, a natural explanation is that they are hadronic molecules, which means the building blocks of these exotic states are usual hadrons and they interact with each other by exchanging color neutral forces. So, a central question of researching exotic states is to decide whether an exotic state is a molecular state or an "elementary" state~(for reviews, see for example, Refs.~\cite{Hosaka:XYZ_Theo&Exp,Yao:Review_on_PW}). 

Neural network is a machine learning algorithm which works for regression and classification problems. It has been applied to many fields in physics, like nuclear physics~\cite{Boehnlein:ML_in_nuclear_phys}, high energy physics experiments~\cite{Gianelle:ML&bjet}, as well as hydrodynamics~\cite{Taradiy:ML&fluid_dynamics}. Recently, it also been applied to high energy physics phenomenology to study~\cite{JPAC:2021rxu,Ng:2021ibr}, for example, intermediate states in $\pi N$ scattering~\cite{Hosaka_piN}, $NN$ scattering~\cite{Hosaka_NN} as well as extracting scattering length and effective range of exotic hadron states~\cite{Liu:ML&Exotic_Hadrons}. Inspired by these applications, an idea is using neural networks to identify hadronic molecular states. 

In this work, the neural networks are trained by the invariant mass spectra generated artificially with labels "0" for molecular states and "1" for elementary states. During training, the validation tests will be done at every training epoch to monitor the network performances and to avoid overfitting. After training, the invariant mass spectrum from real experimental data will be put into trained neural network and the output which is a number from 0 to 1 will describe the possibility to be an elementary state. That means, if the output is closer to 0, then the resonance is more like a molecular state. On the contrary, if the output is closer to 1, then the resonance is more like an elementary state.  

In 2003,  $X(3872)$ was firstly detected by Belle~\cite{Belle_X3872} in the  $J/\psi \pi^{+}\pi^{-}$ final state. Then, it was also observed near the $\bar{D}^{*}D$ threshold~\cite{BarBar_X3872}. 
Similar situation happened on $Z_c(3900)$, which is both observed in the final state of $J/\psi \pi$~\cite{Zc_Jpsipi} and $\bar{D}^{*}D$~\cite{Zc_DD}. So, it is worthwhile to ask whether or not these two exotic states are hadronic molecules of $\bar{D}^{*}D$. After $X(3872)$ and $Z_c(3900)$ observed, many experiments are done to study their properties~\cite{X3872_2013lhc,X3872_2020lhc,Zc0_Jpsipi}. Furthermore, $X(4260)$ has been detected in the $J/\psi \pi \pi$~\cite{X4260_jpsipipi} and $\chi_{c0}\omega$~\cite{X4260_FS2} final state. Since it is near the threshold of $\omega \chi_{c0}$, a valuable question is that wither it is a molecule of $\omega\chi_{c0}$. All these three states have data in at least two channels, especially in the final state whose threshold is near the resonance. So, in this work, the trained neural networks are applied on these three hadronic states.  

This paper is organized as follows, In Sec.~\ref{data_generation}, the method of training data generation and labelling will be introduced. In Sec.~\ref{model_structure}, the structure of neural network classifier is discussed. The results of training, validation tests, as well as applications are displayed in Sec.~\ref{APP}. Finally, some conclusions and outlooks are given. 

\section{Training data generation}\label{data_generation}
A neural network should be trained by many data with labels, means that people should not only put lineshapes of resonances into neural network, but also know every lineshape in the training data standing for a molecular state or an elementary state. So, criteria for the nature of resonances are needed. The pole counting rule~(PCR)~\cite{Morgan_PCR} is a convenient choice. According to PCR, the nature of an S-wave resonance is connected to the number of poles near threshold. If there is only one pole near the threshold of the couple channel, the resonance is a molecular state and if there are two poles near the threshold, the resonance is elementary. This method has been used in many works to study the nature of exotic hadron states~\cite{ZhangO_X3872:,CaoX4660,GongZc3900,Cao_X6900,Chen_X2900}. Then, it demands one to choose a parametrization for resonance amplitudes, in which the parameters can be adjusted easily and then the values of parameters will make difference in pole positions as well as the lineshapes. The Flatt\'e-like parametrization is chosen in this work to generate training data. If two final states are considered, the parametrizations of invariant mass spectra are parametrized as,
\begin{equation}\label{Events}
    \begin{aligned}
    \frac{d\sigma_i}{d\sqrt{s}}&\sim \frac{1}{\sqrt{2\pi}\Delta}\int_{\sqrt{s}-3\Delta}^{\sqrt{s}+3\Delta}\mathcal{R}_{i}(s^{\prime})e^{-\frac{(\sqrt{s}-\sqrt{s^{\prime}})^2}{2\Delta^2}} d\sqrt{s^{\prime}},\\
    \mathcal{R}_i(s) &= \rho_i(s)\left|\frac{e^{i\phi_i}}{s-M^2+iM(g_1\rho_1(s)+g_2\rho_2(s))}+bg_i\right|^2\\
    &+\rho_i(s)(a_i\sqrt{s}+b_i),\\
    \end{aligned}
\end{equation}
where $i=1,2$, $M$ is the bare mass for the resonance, $bg$ is coherent background contributions which takes the formula of first order polynomial of $\sqrt{s}$ and $\rho_i(s)(a_i\sqrt{s}+b_i)$ is the incoherent background. $\phi_{1,2}$ are coherent angles. $\rho_1(s),~\rho_2(s)$ represent two-body phase space factor for final state 1~(FS1) and final state 2~(FS2), respectively. The FS1 stands for lower couple channel whose threshold is much lower than the resonance mass while FS2 stands for the couple channel whose threshold is near the resonance. It means that if the resonance is a hadronic molecule, it should be the composite of the two hadrons in FS2. As an example, for $Z_c(3900)$, the FS1 stands for $J/\psi \pi$ channel and FS2 stands for $\bar{D}^{*} D$ channel. In Eq.~(\ref{Events}), a Gaussian convolution is added to smooth the line-shapes of the resonances, where $\Delta$ is fixed at 3~MeV~\cite{X3872_2020lhc}. This is important for $X(3872)$ and unimportant for $Z_c(3900)$ and $X(4260)$, since the latter have much larger widths.

\begin{figure*}[htpb]
    \centering
    \subfigure[]{
    \includegraphics[scale=0.30]{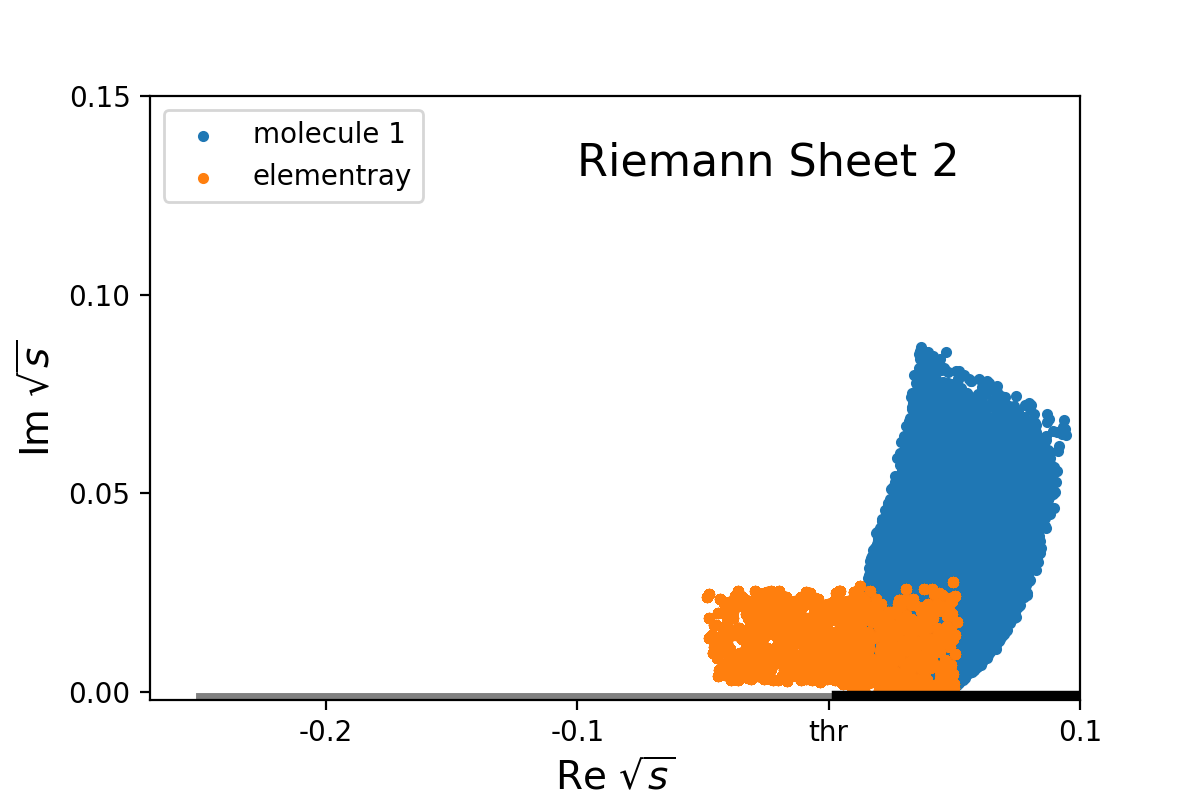}}
    \subfigure[]{
    \includegraphics[scale=0.30]{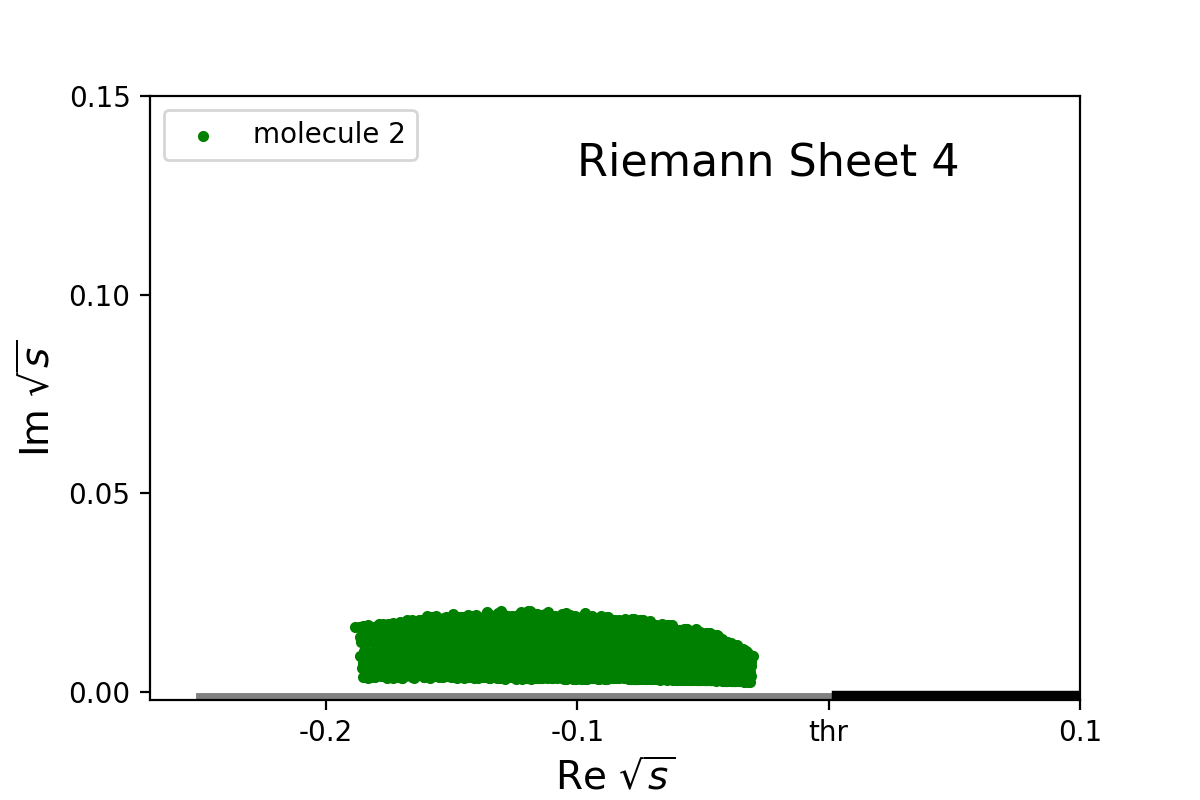}}
    \subfigure[]{
    \includegraphics[scale=0.30]{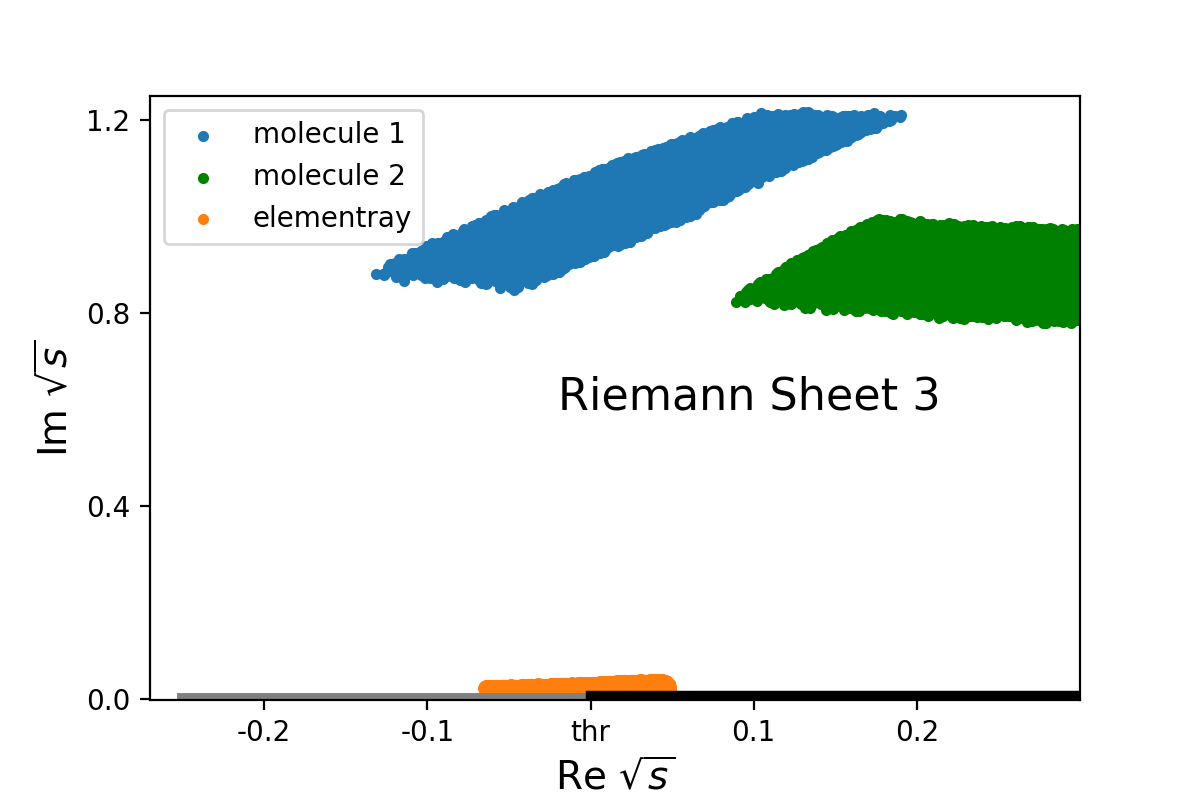}}
    \caption{Pole positions for "elementary" states and molecular states in different Riemann sheets. Here "molecule 1" means a molecular bound state and "molecule 2" means a molecular virtual state.}
    \label{fig:Pole_posi}
\end{figure*}

In practice, the coupling constant $g_1,~g_2$ will be adjusted and then lineshapes as well as the pole positions in the complex $s$ plane will change. Different sheets of complex $s$ plane are defined by changing signs of phase space factors, as shown in Tab.~\ref{tab:RS}. Generally speaking, an elementary state means it hardly couples to FS1 nor FS2, so one should keep $g_1,~g_2$ small to ensure there are two poles near the threshold of FS2. A molecular state indicates that it strongly couples to the second channels, so one should make $g_2 \gg g_1$ to ensure that only one pole is near the threshold of FS2. According to PCR, if the distance between two poles is larger than $\sim$ 200~MeV, one can say there is only one nearby pole~\cite{Chen:PCR&XYZ_states}. The pole positions of training data are shown in Fig~\ref{fig:Pole_posi}.

\begin{table}[htpb]
    \centering
    \begin{tabular}{ccccc}
\hline \hline
& $\mathrm{I}$ & $\mathrm{II}$ & $\mathrm{III}$ & $\mathrm{IV}$ \\
\hline
$\rho_{ 1}$ & $+$ & $-$ & $-$ & $+$ \\
$\rho_{ 2}$ & $+$ & $+$ & $-$ & $-$ \\
\hline 
\hline
\end{tabular}
    \caption{Definition of Riemann sheets}
    \label{tab:RS}
\end{table}

Since the parameter are determined, the training data can be generated using  Eq.~(\ref{Events}). Firstly, one should choose a proper energy range and it can cover the widths of most exotic hadron states. In this paper, the data of FS1 are from $(E_{th2}-100)$~MeV to $(E_{th2}+100)$~MeV and the data of FS2 are from $E_{th2}$ to $(E_{th2}+100)$~MeV, where $E_{thi},~i=1,~2$ are the threshold energies of FS1 or FS2, respectively. It is stressed that the absolute value of $E_{th}$ is unimportant since only the lineshapes will be sent into neural network. 
The energy resolution is fixed at 1~MeV so it can meet the demand that all the inputs sent to a special neural network should have the same size. Furthermore, one can change the energy window sizes in FS1 and FS2, but it can be shown that in most cases, the model can be well trained if the data contain the complete characteristics~(the peaks). The choice of energy window sizes in this work comes from the consideration that these energy regions can cover many hadronic resonances since it will make the models more general rather than only focus on a special resonance. However, in the case of $X(3872)$, it is found that the energy window size for FS2~($\bar{D}^* D$ channel) cannot be taken too small~(e.g., 50~MeV), otherwise the results become unstable because the noises from the data become influential here.

In order to simulate experimental data, the effects of error bars are taken into consideration. The values at every energy point calculated from Eq.~(\ref{Events}) will be taken as the average value of Gaussian distribution, and 5, 10, 15 percent of their values will be taken as the standard deviations~\footnote{The values of standard deviation can also be changed as long as they are not too big to affect the line-shapes seriously.}. Then the Gaussian sampling with these average values and standard deviations gives the data in which the error bar effects are taken into account. At last, the normalization:
\begin{equation}\label{normalize}
    \{ p_i \in \Omega_{\text{training}}\} = \frac{\{ p_i \in \Omega_{\text{training}}\}}{\text{max}\{ p_i \in \Omega_{\text{training}}\}},
\end{equation}
makes points in every sample ranging from 0 to 1, where $\Omega_{\text{training}}$ is the set of points in one sample.

Using the methods introduced above, $4\times 10^4$ groups of labeled invariant mass spectra are generated in total, including $2\times 10^4$ elementary state with label "1" and $2 \times 10^4$ molecular states with label "0". Each group consists of one invariant mass spectrum in FS1 and one invariant mass spectrum in FS2. In practice, $2.4\times 10^4$ groups of invariant mass spectra are used to train the neural network and $1.6\times 10^4$ groups are used to do validation test. In both training set and testing set, the number of elementary states and molecular states are the same.

\section{Construction of neural network model}\label{model_structure}
The goal of the neural network is to establish a map between the input space of invariant mass spectra and the output elementary-molecular classification space. A typical neural network consists of an input layer, some hidden layers and an output layer. In this work, the units in the input layer are numerical values of invariant mass spectra in two final states. Once the inputs are given, they will be linearly transformed with some weights and biases. Then the transformed values will be sent into the activation function, which takes the form of ReLU~(rectified linear unit),
\begin{equation}
    \text{ReLU}(x)=\text{max}(0,x),
\end{equation}
for units in hidden layers, and for units in output layer, it takes the form of sigmoid,
\begin{equation}\label{sigmoid}
    \sigma(x) = \frac{1}{1+e^{-x}}.
\end{equation}
The values passed through activation functions are fed to the units in next layer until meet the output layer, i.e., every unit value is obtained by combining all units in the previous layer linearly and passing through the activation function.
The reason for choosing Eq.~(\ref{sigmoid}) as the activation function for output layer is that it can map arbitary value into a number ranging from 0 to 1, which represents the possibility of being an elementary state.  

Once the output is given, which means the network gives the prediction of the nature of input sample, the discrepancy between the label and the prediction will be calculated using loss function. In this problem, the loss function takes the form of,
\begin{equation}
    \text{loss} = -\frac{1}{n}\sum_{i=1}^{n}\left[q_i\log p_i+(1-q_i)\log(1-p_i)\right],
\end{equation}
where $n$ is the batch size which means the number of samples in each input and $p_i$ is the output for $i$th sample, $q_i$ is the label of $i$th sample in the training batch. 

During the training, the weights and biases used to do the linear combination of unit values should be optimized to have more accurate predictions or smaller loss. Usually, the upgrading of weights and biases use gradient descent algorithm which has been integrated into many packages. The optimzer Adam~\footnote{See: https://pytorch.org} is used in this work. The process from the input to output called forward pass while the upgrading of weights and biases called backpropagation. With circulation of these two processes, the classifier will be trained accurate enough. All these algorithm are performed with PyTorch~\footnote{https://pytorch.org} and more details about neural network can be found in, e.g., Ref.~\cite{Element_SL}.
\begin{figure}[htpb]
    \centering
    \includegraphics[scale=0.33]{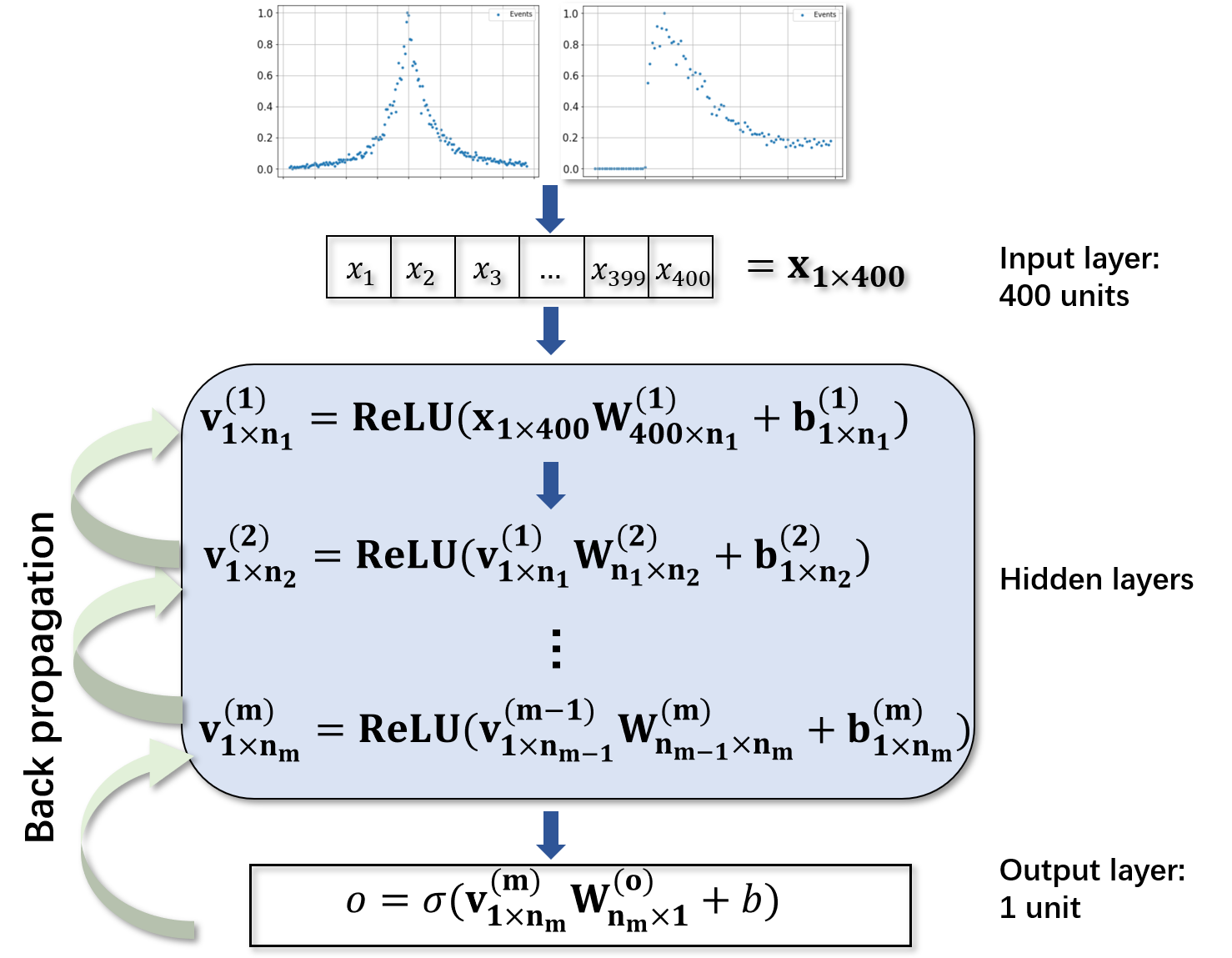}
    \caption{The process of a sample $\mathbf{x}$ crossing the neural network. }
    \label{fig:Network}
\end{figure}

Generally speaking, more hidden layers or hidden layers with more units means there are more parameters including weights and biases involved in the network, so that it can give more accurate predictions with respect to the training data. But on the other hand, if there are too many parameters, the problem of overfitting will arise which means that the network has poor predictive power when it is applied to new data. So, there exist a balance between accuracy and extrapolation. In this work, neural networks with different number of hidden layers and different number of units in these layers are constructed. 

The process of a sample crossing the neural network is displayed in Fig.~(\ref{fig:Network}). The sample is consisted of two invariant mass spectrum and is represented by a vector. The weight matrix and the bias vector are donated as $\mathbf{W},~\mathbf{b}$, respectively. In Fig.~(\ref{fig:Network}), the subscription is the shape of this matrix where $\mathrm{n}_{\mathrm{i}}$ is the number of units in the i-th hidden-layer~(i=1,2,...,m). The superscription (i) represents that this matrix belongs to the i-th hidden-layer and the superscription (o) means that this matrix belongs to the output layer. The output $o$ is a number ranging from 0 to 1.

\section{Training, validation tests and applications}\label{APP}
\subsection{Training and validation tests}
\begin{figure}[htpb]
    \centering
    \includegraphics[scale=0.4]{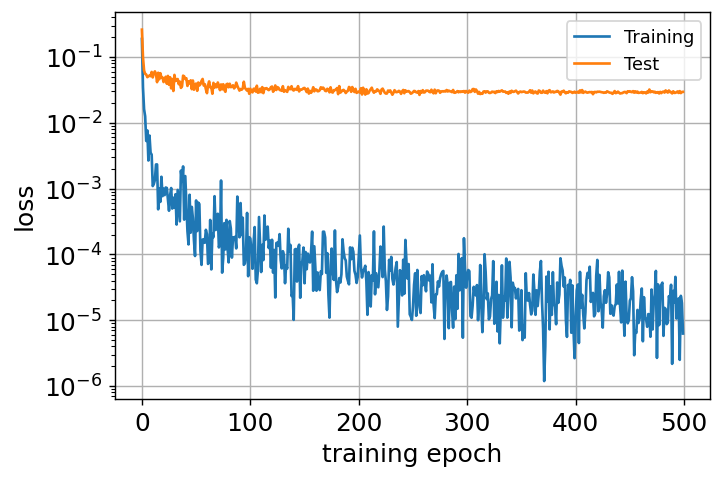}
    \caption{The loss of training and test data.}
    \label{fig:loss}
\end{figure}

\begin{figure}[htpb]
    \centering
    \includegraphics[scale=0.4]{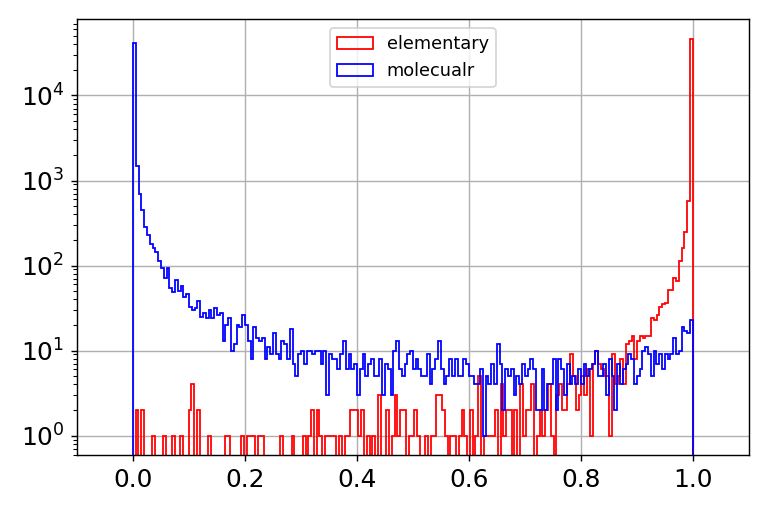}
    \caption{Output distributions of test data.}
    \label{fig:score}
\end{figure}

A model with structure 400-[10-10]-1 is trained, which means its input layer has 400 units, output layer has 1 unit and the two hidden layers both have 10 units.
As mentioned in Sec.~\ref{data_generation}, the performances during training are monitored by validation tests. The data in validation test are different from those in training, but they are generated by the same method. Fig.~\ref{fig:loss} shows the loss of training data as well as test data of the model. It is shown that after 500 training epochs, the losses in both training data as well as test data become small. The training loss usually becomes smaller with more training epochs, but it does not mean that the network will perform better and better with as many training epochs as one wants. If the test loss arrives a small value but become larger with more training epoch, then the neural network is thought to be well trained and more training epochs will cause overfitting, as shown in Fig.~\ref{fig:loss}. It looks like converging to a local minimum of parameter space in conventional fitting. So, the network with the minimum test loss will be picked up.  Further more, the output distribution of samples in test data is shown in Fig.~\ref{fig:score}. From the figure, it is shown that the elementary states labeled by 1 and the molecular states labeled by 0 are classified well since almost all outputs for elementary states are near 1 and for molecular states are near 0.

 \subsection{Applications}

After training and model selection, the trained models is applied to experimental data of $X(3872)$, $X(4260)$ as well as $Z_c(3900)$. For $Z_c(3900)$, the data observed in the $J/\psi \pi$ and $\bar{D}^{*}D$ final states from Ref.~\cite{Zc_Jpsipi} and Ref.~\cite{Zc_DD}, respectively are used. As for $X(3872)$, the data observed in the $J/\psi \pi \pi$ and $\bar{D}^{*}D$ final states from Ref.~\cite{X3872_2020lhc} and Ref.~\cite{BarBar_X3872} are used. For $X(4260)$, the data observed in $J/\psi \pi\pi$~\cite{X4260_jpsipipi} and $\chi_{c0}\omega$~\cite{X4260_FS2} final states are used.
\begin{figure*}[htpb]
    \centering
    \subfigure[]{
    \includegraphics[scale=0.7]{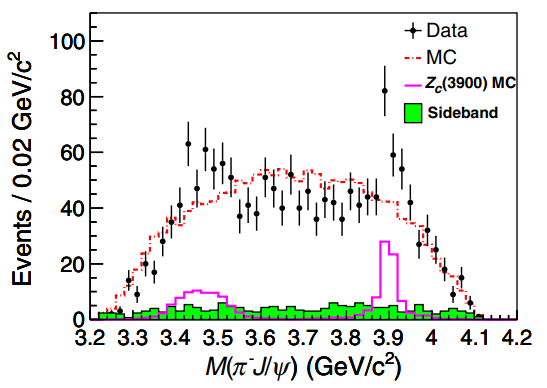}}
    \subfigure[]{
    \includegraphics[scale=0.3]{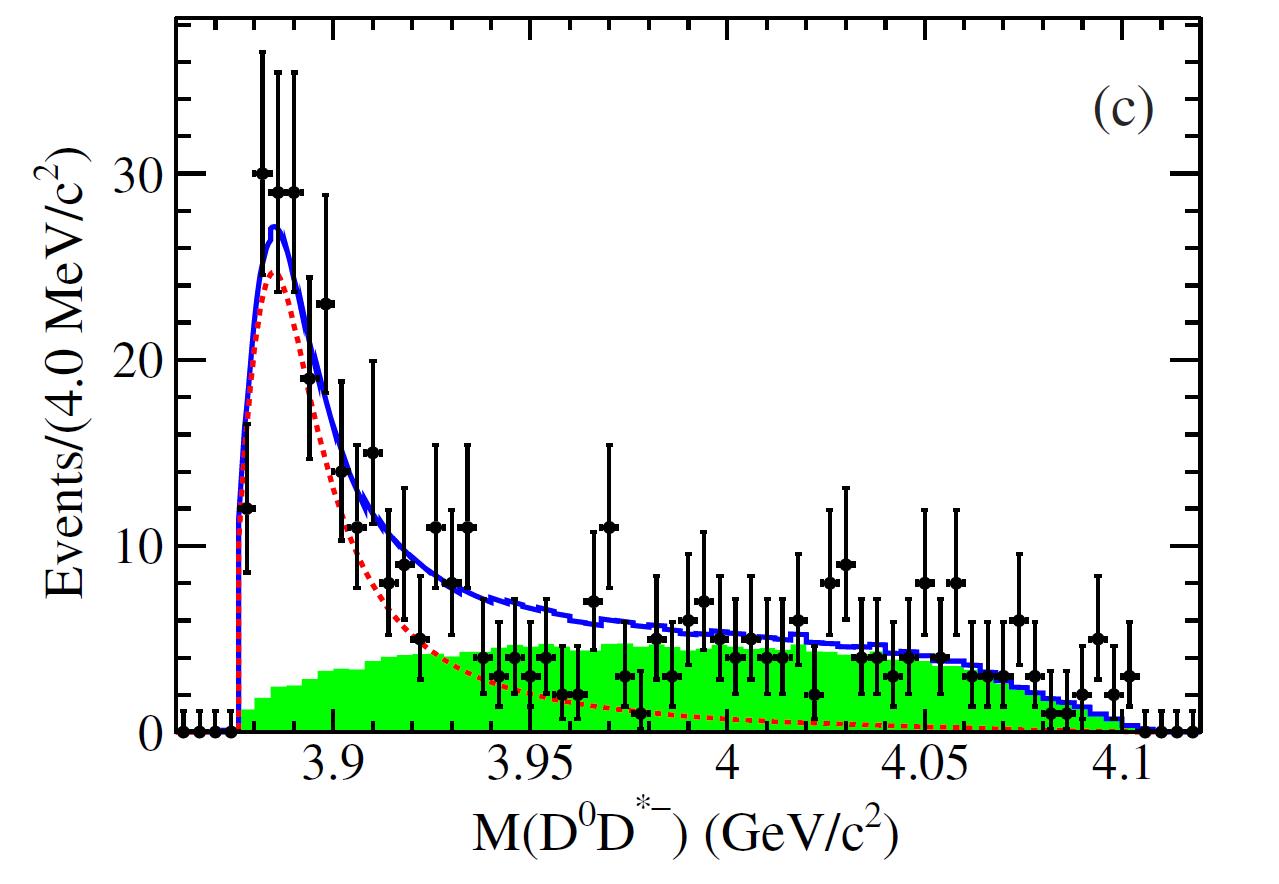}}
    \subfigure[]{
    \includegraphics[scale=0.22]{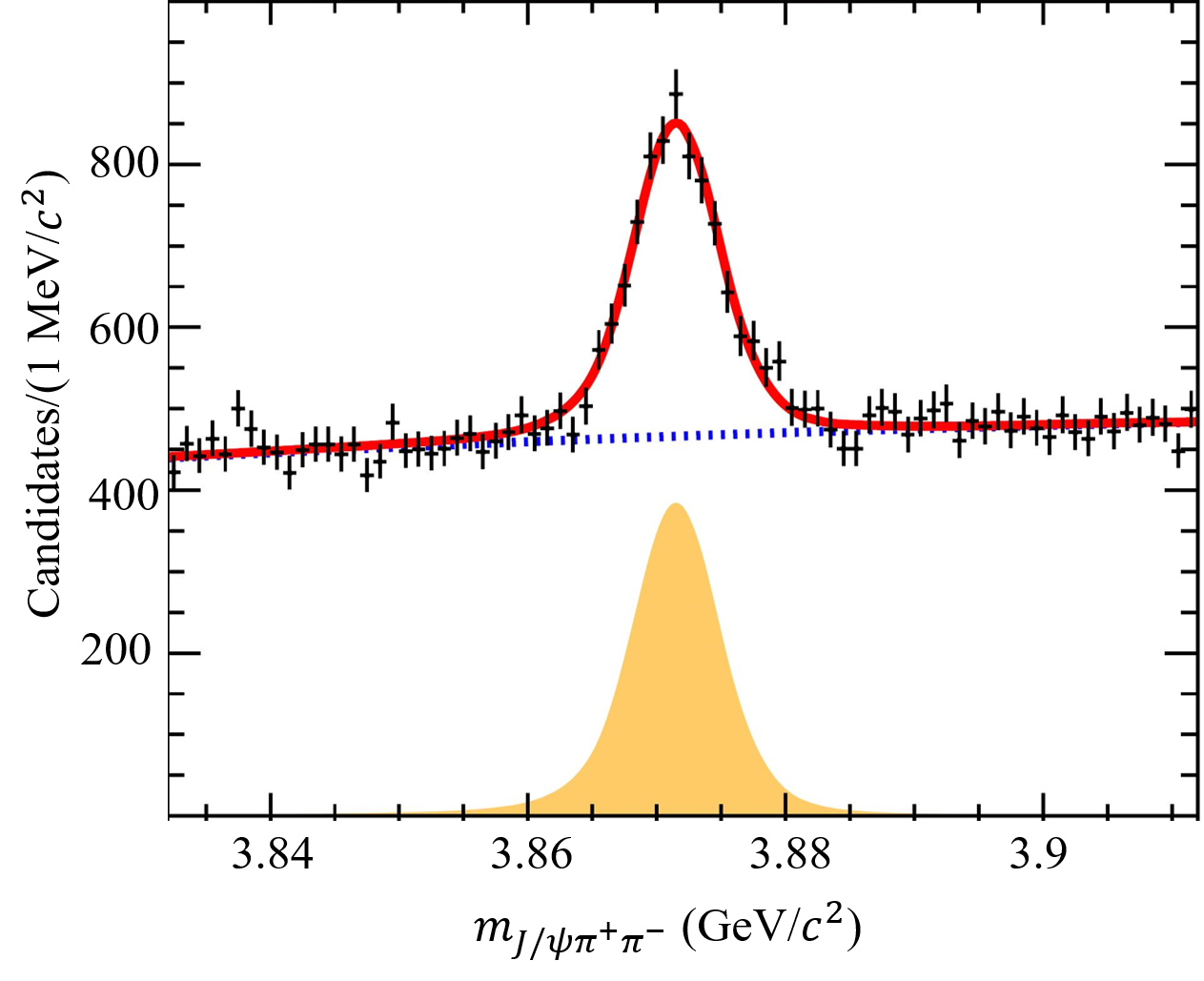}}
    \subfigure[]{
    \includegraphics[scale=0.25]{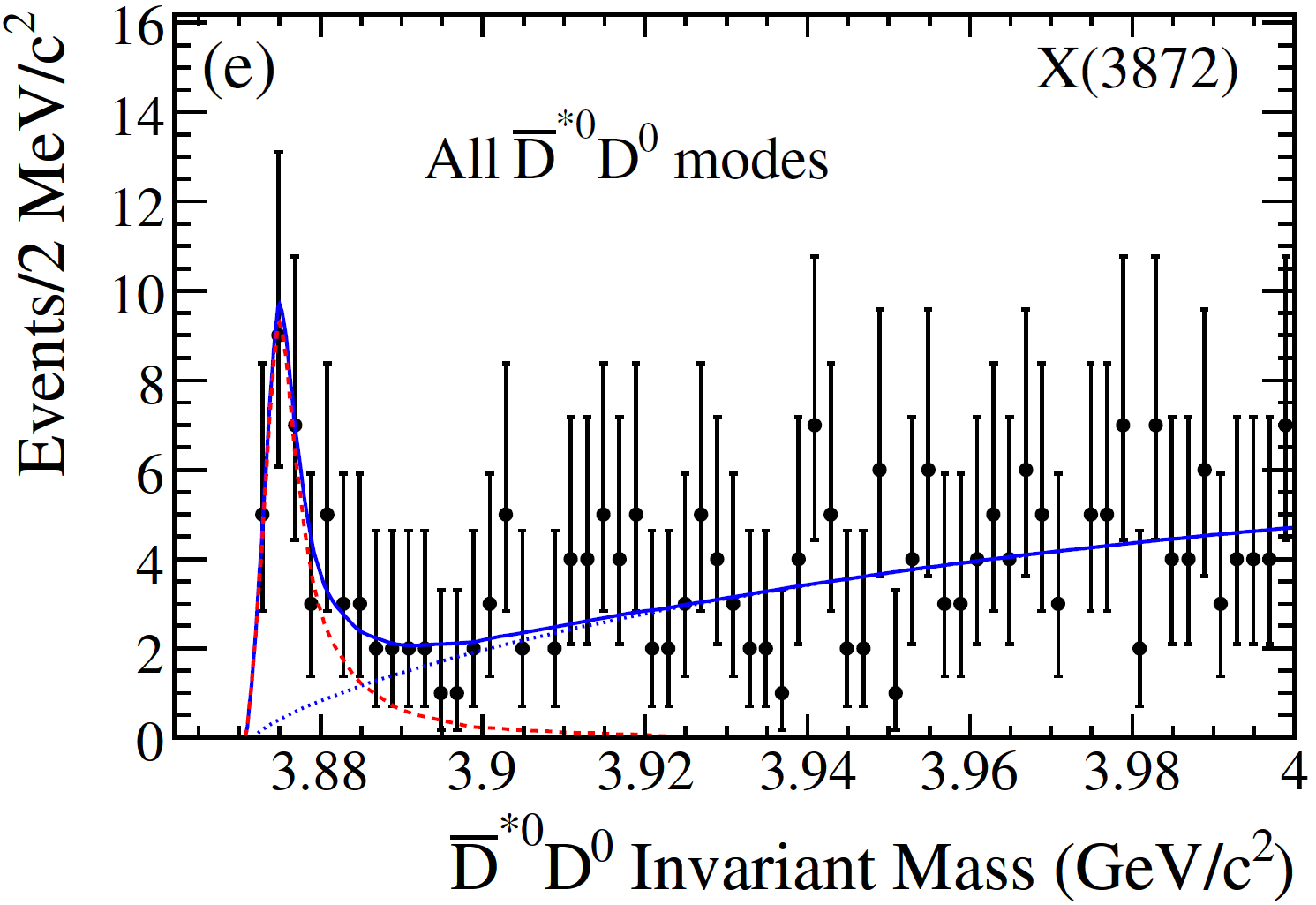}}
    \subfigure[]{
    \includegraphics[scale=0.4]{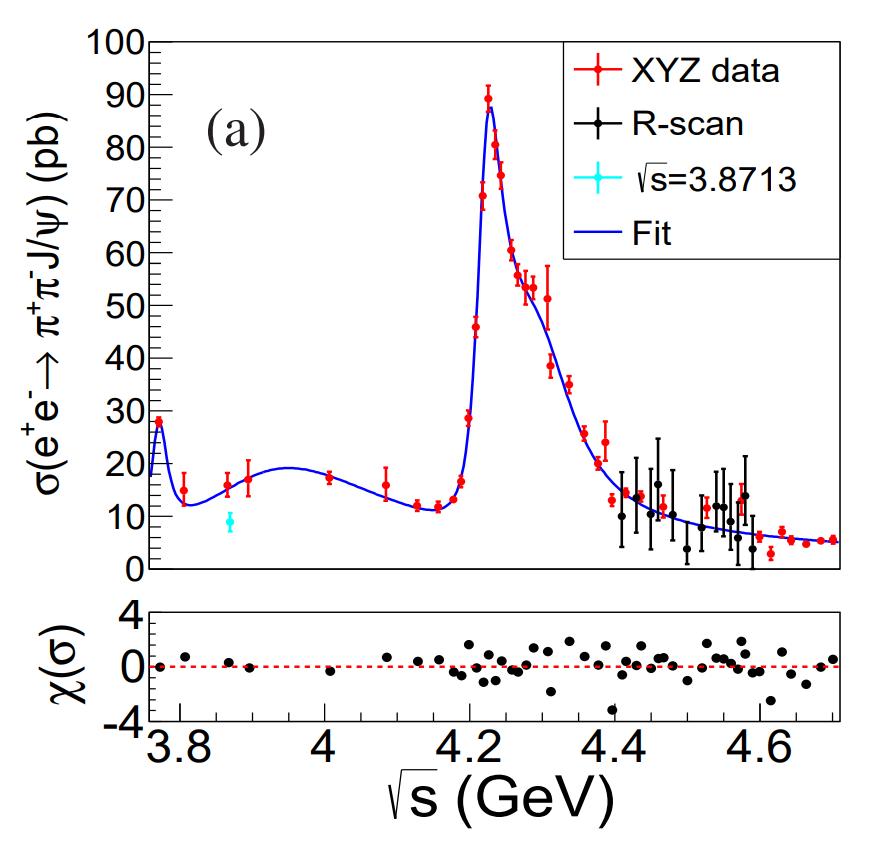}}
    \subfigure[]{
    \includegraphics[scale=0.20]{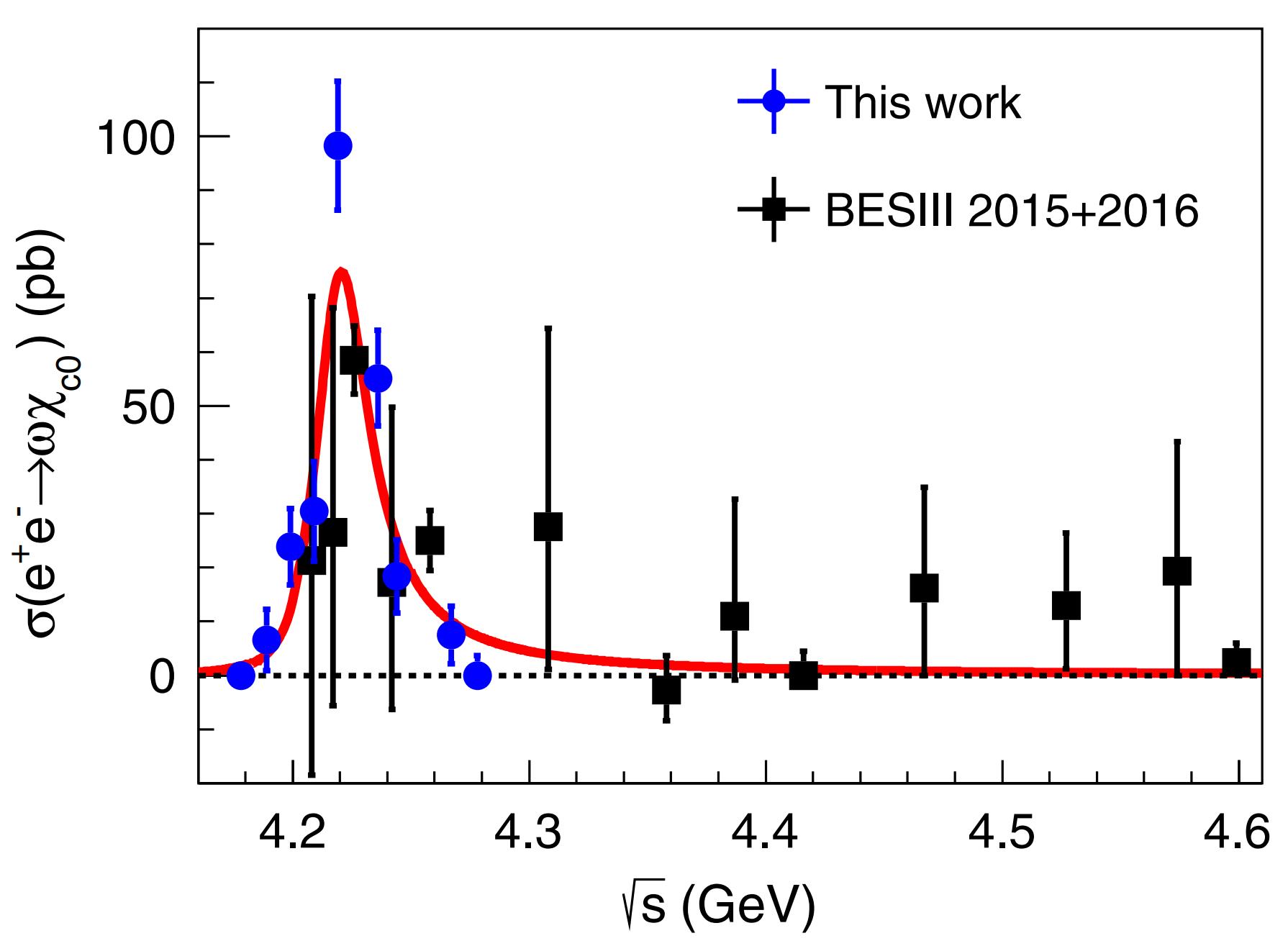}}
    \caption{Experimental data used in the application. (a): $Z_c(3900)$ data from Ref.~\cite{Zc_Jpsipi}, (b): $Z_c(3900)$ data from Ref.~\cite{Zc_DD}, (c): $X(3872)$ data from Ref.~\cite{X3872_2020lhc}, (d): $X(3872)$ data from Ref.~\cite{BarBar_X3872}, (e): $X(4260)$ data from Ref.~\cite{BESIII:2022X4260}, (f): $X(4260)$ data from Ref.~\cite{X4260_FS2}.}
    \label{fig:data_apply}
\end{figure*}

Before sending the experimental data into the neural network, it is noted that the energy resolution are different among experiments. So, these data should be supplied in order to have the same size as input of the neural network in which the energy resolution is fixed at 1~MeV. To be specific, if the energy resolution of experimental data is larger than 1~MeV~(as shown in Fig.~\ref{fig:data_apply} (a), (b) and (d)), then the method of linear interpolation will be employed to supply extra points. For example, the  energy resolution of Fig.~\ref{fig:data_apply} (a) is 10~MeV, so 9 points will be inserted using linear interpolation between two neighbouring experimental data. Of course, the effects of error bar are also taken into consideration, the final application data are obtained by Gaussian sampling in which the central values of experimental data are regarded as average values and the error bars are regarded as standard deviations. At last, 100 invariant mass spectra are obtained for every group of experimental data using Gaussian sampling and the energy resolution is 1~MeV after linear interpolation. After these procedures, the application data should be normalized into 0 to 1 as done in Eq.~(\ref{normalize}).  

\begin{figure*}[htpb]
    \centering
    \subfigure[$X(3872)$]{
    \includegraphics[scale=0.30]{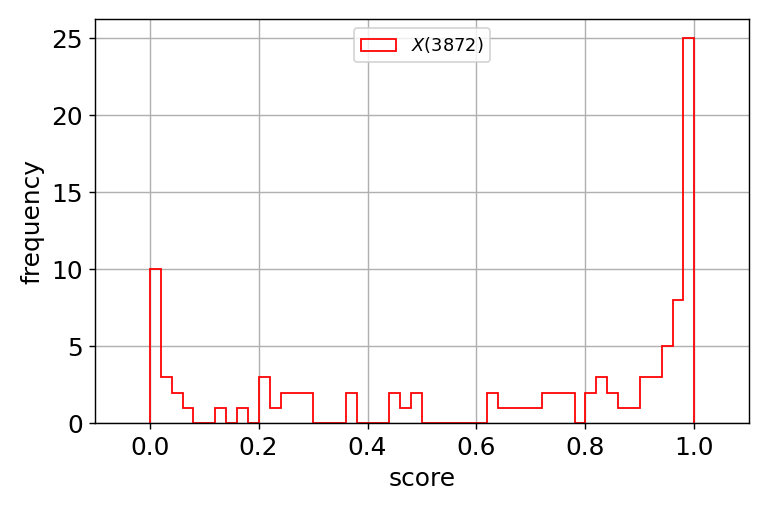}}
    \subfigure[$X4260$]{
    \includegraphics[scale=0.30]{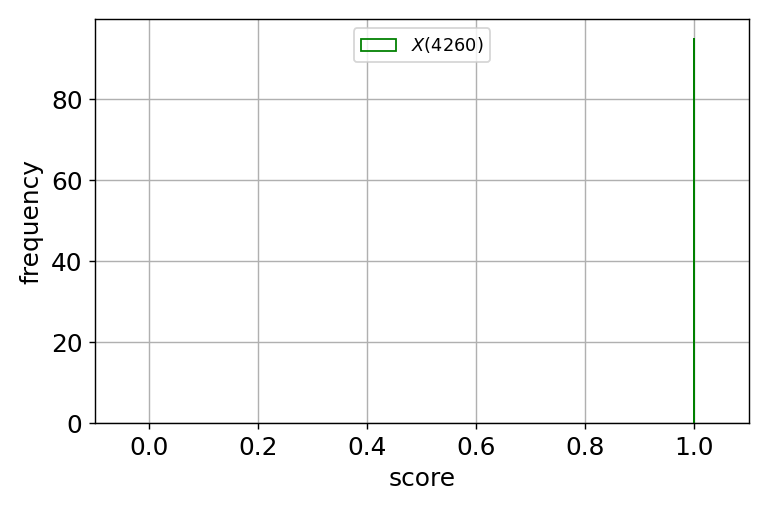}}
    \subfigure[$Z_c(3900)$]{
    \includegraphics[scale=0.30]{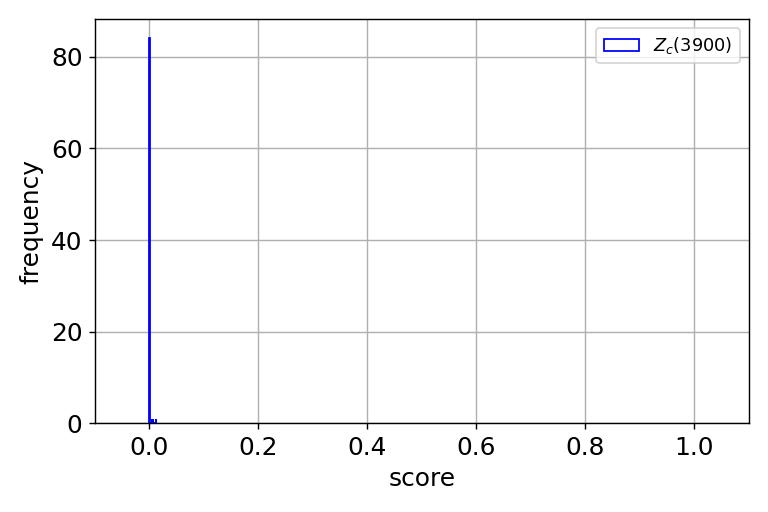}}
    \caption{Output distribution for (a): $X(3872)$, (b): $X(4260)$ and (c): $Z_c(3900)$ using different models.}
    \label{fig:scores}
\end{figure*}

Then, the application data are substituted into trained model to obtain the predictions for the nature of $Z_c(3900)$, $X(4260)$ and $X(3872)$. The predictions for these three states are shown in Fig.~\ref{fig:scores}. 
For $Z_c(3900)$, about all 100 samples in the application data have outputs near 0, means that it strongly couples to $\bar{D}^* D$ channel and should be regarded as a hadronic molecule of $\bar{D}^{*}D$. This interpretation is consistent with Refs.~\cite{GongZc3900,Wilbring_Zc3900,Albaladejo:Zc3900,Karliner_Zc3900}. For $X(3872)$, most outputs of 100 samples are near 1, that means the coupling between $X(3872)$ and $\bar{D}^* D$ channel is not strong enough to dominate its production. In other words, $X(3872)$ behaves more like an elementary state. The neural network prediction for $X(3872)$ is consistent with e.g., Refs.~\cite{ZhangO_X3872:,Ferretti_X3872,Juan_X3872,Tan_X3872}. The predictions for $X(4260)$ are more coincident. All outputs are near 1, which means that $X(4260)$ couples to $\omega \chi_{c0}$ not strong enough and can not be regarded as a molecule of $\omega \chi_{c0}$. This prediction is consistent with e.g., Refs.~\cite{Dai_4260,CaoX4260}.  Furthermore, the measurement of cross section $e^{+} e^{-} \rightarrow \mu^{+} \mu^{-}$ performed by BESIII ~\cite{BESIII:X4260_MuonicWidth} gives the muonic width of $X(4260)$ to be from 1.09 to 1.53~KeV which strongly indicates that $X(4260)$ has a charmonium nature~\cite{CaoX4260}.

Furthermore, other model structures are also constructed. For instance, models with structures 400-[10-5]-1, 400-[15-5-5]-1, 400-[20-20]-1 are trained and the outputs for $X(3872),~Z_c(3900),~X(4260)$ are unchanged.

\begin{figure}[htpb]
    \centering
    \subfigure[]{
    \includegraphics[scale=0.55]{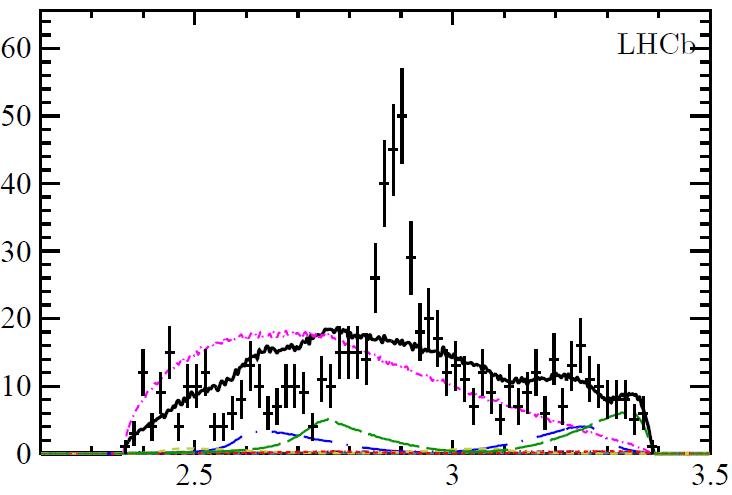}}
    \subfigure[]{
    \includegraphics[scale=0.33]{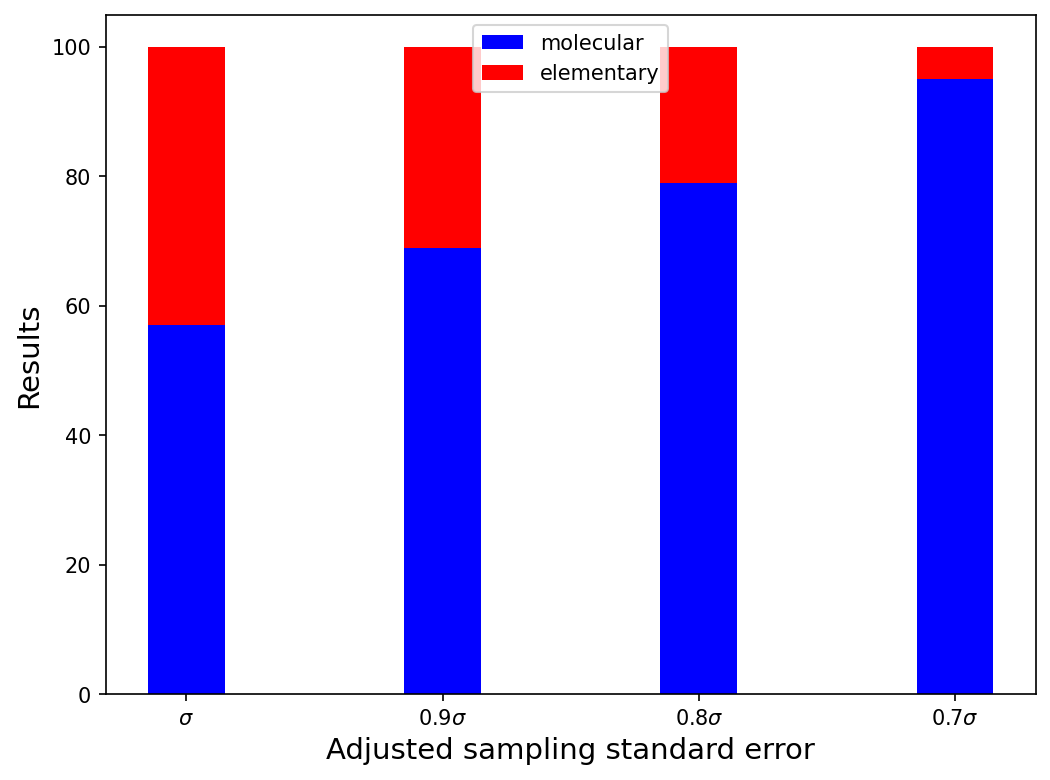}}
    \caption{(a):~Experimental data of $X_1(2900)$~\cite{LHCb:X2900}. (b):~Outputs of $X_1(2900)$. Where the $\sigma$ stands for the experimental error bar.}
    \label{fig:X2900}
\end{figure}

Of course, some exotic hadron states have data only in one final state~(usually only in FS1). At this time, one still can identify whether they are hadronic molecules by neural network. As an example, a network is trained by FS1 data in Eq.~(\ref{Events}) alone. This network is used to identify the properties of $X_1(2900)$, which is regarded as a $\bar{D}_1 K$ molecule~\cite{Chen_X2900}. When the experimental data are sent into network, about a half samples are identified as elementary states and another are molecular states, which means the network is not good enough yet to determine its nature. But this may be caused by the big error bars of the experimental data~(see Fig.\ref{fig:X2900}(a)). When generating samples by Gaussian sampling, their lineshapes can change a lot within error bars which are used as standard deviations. So, one can reduce the standard deviations in Gaussian sampling to avoid misleading. As shown in Fig.~\ref{fig:X2900}(b), with smaller standard deviations, the outputs are converged to a molecular state, i.e., scores of the most samples are $<$ 0.5. That means, neural network can also identify hadronic molecules using only one invariant mass spectrum if the experimental error bars are small enough.

\section{Summary and outlook}
In this work, machine learning models based on neural network are developed and used to decide whether $X(3872)$, $X(4260)$ as well as $Z_c(3900)$ can be regarded as hadronic molecules of special channels. Resonance lineshape data observed in two final states are taken as input, and the output is a number which can be seen as the possibility of this resonance to be an elementary state. In other words, the resonance is more like a hadronic molecule of particles in FS2 whose threshold is near the resonance if the output is closer to 0. The well trained networks are picked up to do these tasks. The results show that $Z_c(3900)$ can be regarded a molecule of $\bar{D}^{*}D$ but $X(3872)$ is not like a molecule of $\bar{D}^{*}D$. Besides, if data of $X(4260)$ in $J/\psi \pi\pi$ and $\omega\chi_{c0}$ channels are taken as input, the predictions of neural networks suggest
that it is not like a molecular state of $\omega\chi_{c0}$\footnote{ It is suggested in Ref.~\cite{X4260_DD1} that $X(4260)$ is a $\bar{D}D_1(2420)$  molecule. Since there is no data in this final state, we cannot study this scenario. Furthermore, Ref.~\cite{CaoX4260} suggested $X(4260)$ can well be described as a $4^3 S_1$ and $3^3 D_1$ mixing state, the molecule picture is no longer appealing.} These interpretations from neural networks are consistent with many previous phenomenological studies. So, the method employed in this work is practicable.

The philosophy of a neural network to do such a binary classification task is to represent the inputs as points in the parameter space and build a hypersurface to separate these points into two classes. The training process is nothing but to adjust the hypersurface to do the separation better. In other words, the output for a set of data reflects that how can the parameters take values to describe such an input. That sounds just like to do a conventional fit. In this work, the data in both two final states perform better compared to data in only one final state, it makes the classification like a joint fitting and the prediction are more reliable.

The method in this work provides a new way to identify the nature of hadron states. It is a data driven, model independent and general method. That means a trained model can be used to classify different resonances in different processes.
In the future, the machine learning method can be developed better to understand the nature of exotic states. For instance, Dalitz plots are more original data compared with invariant mass spectra and convolution neural network is powerful in image recognition. So, it can be used to classify elementary states and hadronic molecules based on Dalitz plots. It is believed that with the development of algorithms and computer technology, machine learning will become more and more popular in the field of particle physics.

\section*{ACKNOWLEDGEMENTS}
This work is supported in part by National Nature Science Foundations of China under Contract Number 11975028 and 10925522.

\newpage
\bibliography{bibfiles}

\begin{thebibliography}{39}
\expandafter\ifx\csname natexlab\endcsname\relax\def\natexlab#1{#1}\fi
\expandafter\ifx\csname bibnamefont\endcsname\relax
  \def\bibnamefont#1{#1}\fi
\expandafter\ifx\csname bibfnamefont\endcsname\relax
  \def\bibfnamefont#1{#1}\fi
\expandafter\ifx\csname citenamefont\endcsname\relax
  \def\citenamefont#1{#1}\fi
\expandafter\ifx\csname url\endcsname\relax
  \def\url#1{\texttt{#1}}\fi
\expandafter\ifx\csname urlprefix\endcsname\relax\def\urlprefix{URL }\fi
\providecommand{\bibinfo}[2]{#2}
\providecommand{\eprint}[2][]{\url{#2}}

\bibitem[{\citenamefont{Hosaka et~al.}(2016)}]{Hosaka:XYZ_Theo&Exp}
\bibinfo{author}{\bibfnamefont{A.}~\bibnamefont{Hosaka}} \bibnamefont{et~al.},
  \bibinfo{journal}{PTEP} \textbf{\bibinfo{volume}{2016}},
  \bibinfo{pages}{062C01} (\bibinfo{year}{2016}), \eprint{arXiv: 1603.09229}.

\bibitem[{\citenamefont{Yao et~al.}(2021)}]{Yao:Review_on_PW}
\bibinfo{author}{\bibfnamefont{D.-L.} \bibnamefont{Yao}} \bibnamefont{et~al.},
  \bibinfo{journal}{Rept. Prog. Phys.} \textbf{\bibinfo{volume}{84}},
  \bibinfo{pages}{076201} (\bibinfo{year}{2021}), \eprint{arXiv: 2009.13495}.

\bibitem[{\citenamefont{Boehnlein et~al.}(2021)}]{Boehnlein:ML_in_nuclear_phys}
\bibinfo{author}{\bibfnamefont{A.}~\bibnamefont{Boehnlein}}
  \bibnamefont{et~al.} (\bibinfo{year}{2021}), \eprint{arXiv: 2112.02309}.

\bibitem[{\citenamefont{Gianelle et~al.}(2022)}]{Gianelle:ML&bjet}
\bibinfo{author}{\bibfnamefont{A.}~\bibnamefont{Gianelle}} \bibnamefont{et~al.}
  (\bibinfo{year}{2022}), \eprint{arXiv: 2202.13943}.

\bibitem[{\citenamefont{Taradiy et~al.}(2021)}]{Taradiy:ML&fluid_dynamics}
\bibinfo{author}{\bibfnamefont{K.}~\bibnamefont{Taradiy}} \bibnamefont{et~al.}
  (\bibinfo{year}{2021}), \eprint{arXiv: 2106.02841}.

\bibitem[{\citenamefont{Albaladejo et~al.}(2021)}]{JPAC:2021rxu}
\bibinfo{author}{\bibfnamefont{M.}~\bibnamefont{Albaladejo}}
  \bibnamefont{et~al.} (\bibinfo{collaboration}{JPAC}) (\bibinfo{year}{2021}),
  \eprint{arXiv:2112.13436}.

\bibitem[{\citenamefont{Ng et~al.}(2022)}]{Ng:2021ibr}
\bibinfo{author}{\bibfnamefont{L.}~\bibnamefont{Ng}} \bibnamefont{et~al.}
  (\bibinfo{collaboration}{JPAC}), \bibinfo{journal}{Phys. Rev. D}
  \textbf{\bibinfo{volume}{105}}, \bibinfo{pages}{L091501}
  (\bibinfo{year}{2022}), \eprint{arXiv:2110.13742}.

\bibitem[{\citenamefont{Sombillo et~al.}(2021{\natexlab{a}})}]{Hosaka_piN}
\bibinfo{author}{\bibfnamefont{D.}~\bibnamefont{Sombillo}}
  \bibnamefont{et~al.}, \bibinfo{journal}{Phys. Rev. D}
  \textbf{\bibinfo{volume}{104}} (\bibinfo{year}{2021}{\natexlab{a}}), ISSN
  \bibinfo{issn}{2470-0029},
  \urlprefix\url{http://dx.doi.org/10.1103/PhysRevD.104.036001}.

\bibitem[{\citenamefont{Sombillo et~al.}(2021{\natexlab{b}})}]{Hosaka_NN}
\bibinfo{author}{\bibfnamefont{D.}~\bibnamefont{Sombillo}}
  \bibnamefont{et~al.}, \bibinfo{journal}{Few-Body Syst.}
  \textbf{\bibinfo{volume}{62}} (\bibinfo{year}{2021}{\natexlab{b}}), ISSN
  \bibinfo{issn}{1432-5411},
  \urlprefix\url{http://dx.doi.org/10.1007/s00601-021-01642-z}.

\bibitem[{\citenamefont{Liu et~al.}(2022)}]{Liu:ML&Exotic_Hadrons}
\bibinfo{author}{\bibfnamefont{J.}~\bibnamefont{Liu}} \bibnamefont{et~al.}
  (\bibinfo{year}{2022}), \eprint{arXiv:2202.04929}.

\bibitem[{\citenamefont{Choi et~al.}(2003)}]{Belle_X3872}
\bibinfo{author}{\bibfnamefont{S.-K.} \bibnamefont{Choi}} \bibnamefont{et~al.}
  (\bibinfo{collaboration}{Belle Collaboration}), \bibinfo{journal}{Phys. Rev.
  Lett.} \textbf{\bibinfo{volume}{91}}, \bibinfo{pages}{262001}
  (\bibinfo{year}{2003}),
  \urlprefix\url{https://link.aps.org/doi/10.1103/PhysRevLett.91.262001}.

\bibitem[{\citenamefont{Aubert et~al.}(2008)}]{BarBar_X3872}
\bibinfo{author}{\bibfnamefont{B.}~\bibnamefont{Aubert}} \bibnamefont{et~al.}
  (\bibinfo{collaboration}{BABAR Collaboration}), \bibinfo{journal}{Phys. Rev.
  D} \textbf{\bibinfo{volume}{77}}, \bibinfo{pages}{011102}
  (\bibinfo{year}{2008}),
  \urlprefix\url{https://link.aps.org/doi/10.1103/PhysRevD.77.011102}.

\bibitem[{\citenamefont{Ablikim et~al.}(2013)}]{Zc_Jpsipi}
\bibinfo{author}{\bibfnamefont{M.}~\bibnamefont{Ablikim}} \bibnamefont{et~al.}
  (\bibinfo{collaboration}{BESIII Collaboration}), \bibinfo{journal}{Phys. Rev.
  Lett.} \textbf{\bibinfo{volume}{110}}, \bibinfo{pages}{252001}
  (\bibinfo{year}{2013}),
  \urlprefix\url{https://link.aps.org/doi/10.1103/PhysRevLett.110.252001}.

\bibitem[{\citenamefont{Ablikim et~al.}(2015{\natexlab{a}})}]{Zc_DD}
\bibinfo{author}{\bibfnamefont{M.}~\bibnamefont{Ablikim}} \bibnamefont{et~al.}
  (\bibinfo{collaboration}{BESIII Collaboration}), \bibinfo{journal}{Phys. Rev.
  D} \textbf{\bibinfo{volume}{92}}, \bibinfo{pages}{092006}
  (\bibinfo{year}{2015}{\natexlab{a}}),
  \urlprefix\url{https://link.aps.org/doi/10.1103/PhysRevD.92.092006}.

\bibitem[{\citenamefont{Aaij et~al.}(2013)}]{X3872_2013lhc}
\bibinfo{author}{\bibfnamefont{R.}~\bibnamefont{Aaij}} \bibnamefont{et~al.}
  (\bibinfo{collaboration}{LHCb Collaboration}), \bibinfo{journal}{Phys. Rev.
  Lett.} \textbf{\bibinfo{volume}{110}}, \bibinfo{pages}{222001}
  (\bibinfo{year}{2013}),
  \urlprefix\url{https://link.aps.org/doi/10.1103/PhysRevLett.110.222001}.

\bibitem[{\citenamefont{Aaij et~al.}(2020{\natexlab{a}})}]{X3872_2020lhc}
\bibinfo{author}{\bibfnamefont{R.}~\bibnamefont{Aaij}} \bibnamefont{et~al.}
  (\bibinfo{collaboration}{LHCb Collaboration}), \bibinfo{journal}{Phys. Rev.
  D} \textbf{\bibinfo{volume}{102}}, \bibinfo{pages}{092005}
  (\bibinfo{year}{2020}{\natexlab{a}}),
  \urlprefix\url{https://link.aps.org/doi/10.1103/PhysRevD.102.092005}.

\bibitem[{\citenamefont{Ablikim et~al.}(2015{\natexlab{b}})}]{Zc0_Jpsipi}
\bibinfo{author}{\bibfnamefont{M.}~\bibnamefont{Ablikim}} \bibnamefont{et~al.}
  (\bibinfo{collaboration}{BESIII Collaboration}), \bibinfo{journal}{Phys. Rev.
  Lett.} \textbf{\bibinfo{volume}{115}}, \bibinfo{pages}{112003}
  (\bibinfo{year}{2015}{\natexlab{b}}),
  \urlprefix\url{https://link.aps.org/doi/10.1103/PhysRevLett.115.112003}.

\bibitem[{\citenamefont{Ablikim et~al.}(2017)}]{X4260_jpsipipi}
\bibinfo{author}{\bibfnamefont{M.}~\bibnamefont{Ablikim}} \bibnamefont{et~al.}
  (\bibinfo{collaboration}{BESIII Collaboration}), \bibinfo{journal}{Phys. Rev.
  Lett.} \textbf{\bibinfo{volume}{118}}, \bibinfo{pages}{092001}
  (\bibinfo{year}{2017}),
  \urlprefix\url{https://link.aps.org/doi/10.1103/PhysRevLett.118.092001}.

\bibitem[{\citenamefont{Ablikim et~al.}(2019)}]{X4260_FS2}
\bibinfo{author}{\bibfnamefont{M.}~\bibnamefont{Ablikim}} \bibnamefont{et~al.}
  (\bibinfo{collaboration}{BESIII Collaboration}), \bibinfo{journal}{Phys. Rev.
  D} \textbf{\bibinfo{volume}{99}}, \bibinfo{pages}{091103}
  (\bibinfo{year}{2019}),
  \urlprefix\url{https://link.aps.org/doi/10.1103/PhysRevD.99.091103}.

\bibitem[{\citenamefont{Morgan}(1992)}]{Morgan_PCR}
\bibinfo{author}{\bibfnamefont{D.}~\bibnamefont{Morgan}},
  \bibinfo{journal}{Nucl. Phys. A} \textbf{\bibinfo{volume}{543}},
  \bibinfo{pages}{632} (\bibinfo{year}{1992}).

\bibitem[{\citenamefont{Zhang et~al.}(2009)\citenamefont{Zhang, Meng, and
  Zheng}}]{ZhangO_X3872:}
\bibinfo{author}{\bibfnamefont{O.}~\bibnamefont{Zhang}},
  \bibinfo{author}{\bibfnamefont{C.}~\bibnamefont{Meng}}, \bibnamefont{and}
  \bibinfo{author}{\bibfnamefont{H.~Q.} \bibnamefont{Zheng}},
  \bibinfo{journal}{Phys. Lett. B} \textbf{\bibinfo{volume}{680}},
  \bibinfo{pages}{453} (\bibinfo{year}{2009}).

\bibitem[{\citenamefont{Cao et~al.}(2019)}]{CaoX4660}
\bibinfo{author}{\bibfnamefont{Q.-F.} \bibnamefont{Cao}} \bibnamefont{et~al.},
  \bibinfo{journal}{Phys. Rev. D} \textbf{\bibinfo{volume}{100}},
  \bibinfo{pages}{054040} (\bibinfo{year}{2019}).

\bibitem[{\citenamefont{Gong et~al.}(2016)}]{GongZc3900}
\bibinfo{author}{\bibfnamefont{Q.-R.} \bibnamefont{Gong}} \bibnamefont{et~al.},
  \bibinfo{journal}{Phys. Rev. D} \textbf{\bibinfo{volume}{94}},
  \bibinfo{pages}{114019} (\bibinfo{year}{2016}).

\bibitem[{\citenamefont{Cao et~al.}(2021{\natexlab{a}})}]{Cao_X6900}
\bibinfo{author}{\bibfnamefont{Q.-F.} \bibnamefont{Cao}} \bibnamefont{et~al.},
  \bibinfo{journal}{Chin. Phys. C} \textbf{\bibinfo{volume}{45}},
  \bibinfo{pages}{103102} (\bibinfo{year}{2021}{\natexlab{a}}).

\bibitem[{\citenamefont{Chen et~al.}(2021)\citenamefont{Chen, Qi, and
  Zheng}}]{Chen_X2900}
\bibinfo{author}{\bibfnamefont{H.}~\bibnamefont{Chen}},
  \bibinfo{author}{\bibfnamefont{H.-R.} \bibnamefont{Qi}}, \bibnamefont{and}
  \bibinfo{author}{\bibfnamefont{H.-Q.} \bibnamefont{Zheng}},
  \bibinfo{journal}{Eur. Phys. J. C} \textbf{\bibinfo{volume}{81}},
  \bibinfo{pages}{812} (\bibinfo{year}{2021}).

\bibitem[{\citenamefont{Chen and Qi}(2022)}]{Chen:PCR&XYZ_states}
\bibinfo{author}{\bibfnamefont{H.}~\bibnamefont{Chen}} \bibnamefont{and}
  \bibinfo{author}{\bibfnamefont{H.-R.} \bibnamefont{Qi}}, in
  \emph{\bibinfo{booktitle}{{10th International workshop on Chiral Dynamics}}}
  (\bibinfo{year}{2022}), \eprint{arXiv:2202.10736}.

\bibitem[{\citenamefont{Hastie et~al.}(2009)\citenamefont{Hastie, Tibshirani,
  and Friedman}}]{Element_SL}
\bibinfo{author}{\bibfnamefont{T.}~\bibnamefont{Hastie}},
  \bibinfo{author}{\bibfnamefont{R.}~\bibnamefont{Tibshirani}},
  \bibnamefont{and} \bibinfo{author}{\bibfnamefont{J.}~\bibnamefont{Friedman}},
  \emph{\bibinfo{title}{The Elements of Statistical Learning}}
  (\bibinfo{publisher}{Springer, New York, NY}, \bibinfo{year}{2009}), ISBN
  \bibinfo{isbn}{978-0-387-84858-7}.

\bibitem[{\citenamefont{Ablikim et~al.}(2022)}]{BESIII:2022X4260}
\bibinfo{author}{\bibfnamefont{M.}~\bibnamefont{Ablikim}} \bibnamefont{et~al.}
  (\bibinfo{collaboration}{BESIII}) (\bibinfo{year}{2022}), \eprint{arXiv:
  2206.08554}.

\bibitem[{\citenamefont{Wilbring et~al.}(2013)\citenamefont{Wilbring, Hammer,
  and Meißner}}]{Wilbring_Zc3900}
\bibinfo{author}{\bibfnamefont{E.}~\bibnamefont{Wilbring}},
  \bibinfo{author}{\bibfnamefont{H.-W.} \bibnamefont{Hammer}},
  \bibnamefont{and} \bibinfo{author}{\bibfnamefont{U.-G.}
  \bibnamefont{Meißner}}, \bibinfo{journal}{Phys. Lett. B}
  \textbf{\bibinfo{volume}{726}}, \bibinfo{pages}{326} (\bibinfo{year}{2013}),
  ISSN \bibinfo{issn}{0370-2693}.

\bibitem[{\citenamefont{Albaladejo et~al.}(2016)}]{Albaladejo:Zc3900}
\bibinfo{author}{\bibfnamefont{M.}~\bibnamefont{Albaladejo}}
  \bibnamefont{et~al.}, \bibinfo{journal}{Phys. Lett. B}
  \textbf{\bibinfo{volume}{755}}, \bibinfo{pages}{337} (\bibinfo{year}{2016}).

\bibitem[{\citenamefont{Karliner and Rosner}(2015)}]{Karliner_Zc3900}
\bibinfo{author}{\bibfnamefont{M.}~\bibnamefont{Karliner}} \bibnamefont{and}
  \bibinfo{author}{\bibfnamefont{J.~L.} \bibnamefont{Rosner}},
  \bibinfo{journal}{Phys. Rev. Lett.} \textbf{\bibinfo{volume}{115}}
  (\bibinfo{year}{2015}), ISSN \bibinfo{issn}{1079-7114},
  \urlprefix\url{http://dx.doi.org/10.1103/PhysRevLett.115.122001}.

\bibitem[{\citenamefont{Ferretti et~al.}(2013)\citenamefont{Ferretti, Galat\`a,
  and Santopinto}}]{Ferretti_X3872}
\bibinfo{author}{\bibfnamefont{J.}~\bibnamefont{Ferretti}},
  \bibinfo{author}{\bibfnamefont{G.}~\bibnamefont{Galat\`a}}, \bibnamefont{and}
  \bibinfo{author}{\bibfnamefont{E.}~\bibnamefont{Santopinto}},
  \bibinfo{journal}{Phys. Rev. C} \textbf{\bibinfo{volume}{88}},
  \bibinfo{pages}{015207} (\bibinfo{year}{2013}).

\bibitem[{\citenamefont{Meng et~al.}(2015)}]{Juan_X3872}
\bibinfo{author}{\bibfnamefont{C.}~\bibnamefont{Meng}} \bibnamefont{et~al.},
  \bibinfo{journal}{Phys. Rev. D} \textbf{\bibinfo{volume}{92}},
  \bibinfo{pages}{034020} (\bibinfo{year}{2015}).

\bibitem[{\citenamefont{Tan and Ping}(2019)}]{Tan_X3872}
\bibinfo{author}{\bibfnamefont{Y.}~\bibnamefont{Tan}} \bibnamefont{and}
  \bibinfo{author}{\bibfnamefont{J.-L.} \bibnamefont{Ping}},
  \bibinfo{journal}{Phys. Rev. D} \textbf{\bibinfo{volume}{100}},
  \bibinfo{pages}{034022} (\bibinfo{year}{2019}).

\bibitem[{\citenamefont{Dai et~al.}(2015)}]{Dai_4260}
\bibinfo{author}{\bibfnamefont{L.~Y.} \bibnamefont{Dai}} \bibnamefont{et~al.},
  \bibinfo{journal}{Phys. Rev. D} \textbf{\bibinfo{volume}{92}},
  \bibinfo{pages}{014020} (\bibinfo{year}{2015}),
  \urlprefix\url{https://link.aps.org/doi/10.1103/PhysRevD.92.014020}.

\bibitem[{\citenamefont{Cao et~al.}(2021{\natexlab{b}})}]{CaoX4260}
\bibinfo{author}{\bibfnamefont{Q.-F.} \bibnamefont{Cao}} \bibnamefont{et~al.},
  \bibinfo{journal}{Eur. Phys. J. C} \textbf{\bibinfo{volume}{81}},
  \bibinfo{pages}{83} (\bibinfo{year}{2021}{\natexlab{b}}).

\bibitem[{BES(2020)}]{BESIII:X4260_MuonicWidth}
\bibinfo{journal}{Phys. Rev. D} \textbf{\bibinfo{volume}{102}},
  \bibinfo{pages}{112009} (\bibinfo{year}{2020}), \eprint{arXiv:2007.12872}.

\bibitem[{\citenamefont{Aaij et~al.}(2020{\natexlab{b}})}]{LHCb:X2900}
\bibinfo{author}{\bibfnamefont{R.}~\bibnamefont{Aaij}} \bibnamefont{et~al.}
  (\bibinfo{collaboration}{LHCb}), \bibinfo{journal}{Phys. Rev. D}
  \textbf{\bibinfo{volume}{102}}, \bibinfo{pages}{112003}
  (\bibinfo{year}{2020}{\natexlab{b}}), \eprint{arXiv: 2009.00026}.

\bibitem[{\citenamefont{Cleven et~al.}(2014)}]{X4260_DD1}
\bibinfo{author}{\bibfnamefont{M.}~\bibnamefont{Cleven}} \bibnamefont{et~al.},
  \bibinfo{journal}{Phys. Rev. D} \textbf{\bibinfo{volume}{90}},
  \bibinfo{pages}{074039} (\bibinfo{year}{2014}), \eprint{arXiv:1310.2190}.

\end{thebibliography}
\end{document}